\documentclass[11pt]{article}
\usepackage{moriond,epsfig,amsmath}

\begin{document}
\vspace{-3cm}
\title{Dark Matter: Connecting LHC searches to direct detection }

\author{ANDREAS CRIVELLIN${}^{*1}$, MARTIN HOFERICHTER${}^{2,3}$, MASSIMILIANO PROCURA${}^4$ \\and LEWIS TUNSTALL${}^5$}
\address{${}^*$Speaker\\
${}^1$CERN Theory Division, CH--1211 Geneva 23, Switzerland\\
${}^2$Institut f\"ur Kernphysik, Technische Universit\"at Darmstadt, D--64289 Darmstadt, Germany\\
${}^3$ExtreMe Matter Institute EMMI, GSI Helmholtzzentrum f\"ur Schwerionenforschung GmbH, D--64291 Darmstadt, Germany\\
${}^4$Fakult\"at f\"ur Physik, Universit\"at Wien, Boltzmanngasse 5, A--1090 Vienna, Austria\\
${}^5$Albert Einstein Center for Fundamental Physics, Institute for Theoretical Physics,\\ University of Bern, Sidlerstrasse 5, CH--3012 Bern, Switzerland}

\maketitle
\abstracts{In these proceedings we review the interplay between LHC searches for dark matter and direct detection experiments. For this purpose we consider two prime examples: the effective field theory (EFT) approach and the minimal supersymmetric standard model (MSSM). In the EFT scenario we show that for operators which do not enter directly direct detection at tree-level, but only via loop effects, LHC searches give complementary constraints. In the MSSM stop and Higgs exchange contribute to the direct detection amplitude. Therefore, LHC searches for supersymmetric particles and heavy Higgses place constraints on the same parameter space as direct detection.}

\section{Introduction}
\label{intro}

Establishing the microscopic nature of Dark Matter (DM) is one of the central, open questions in cosmology and particle physics. In the context of cold nonbaryonic DM, the prevailing paradigm is based on weakly interacting massive particles (WIMPs), and extensive theoretical and experimental resources have been devoted towards identifying viable candidates and developing methods to detect them. 

One of the most studied WIMPs scenarios arises in the Minimal Supersymmetric Standard Model (MSSM), where an assumed $R$-parity ensures that the lightest superpartner (LSP) is a stable neutralino $\chi$ composed of bino, wino, and Higgsino eigenstates. The interactions between DM and the SM particles are mainly mediated by squark and Higgses in the case of bino like DM.

However, it is also possible to study DM interactions with the SM particles in a model independent way by using an effective field theory approach in which the particles mediating the interactions are assumed to be heavy and are integrated out. A main strength of this approach is to provide model-independent relations among distinct null DM searches~\cite{Goodman:2010yf}. As different search strategies probe different energy scales, this separation of scales can have striking consequences when a connection between direct detection experiments and LHC searches is done.

\section{Effective Field Theory}

For operators contributing directly to spin independent scattering, direct detection gives in general much better constraints than LHC searches. As was shown in Ref.~\cite{Frandsen:2012db,Crivellin:2014qxa,Crivellin:2014gpa,D'Eramo:2014aba} there are cases in which operators which do not contribute to spin independent scattering at tree-level, but enter at the one-loop level. As in this case direct detection is loop suppressed, LHC searches can give competitive and complementary constraints.

\begin{figure}[t]
	\centering
    \includegraphics[width=0.48\textwidth]{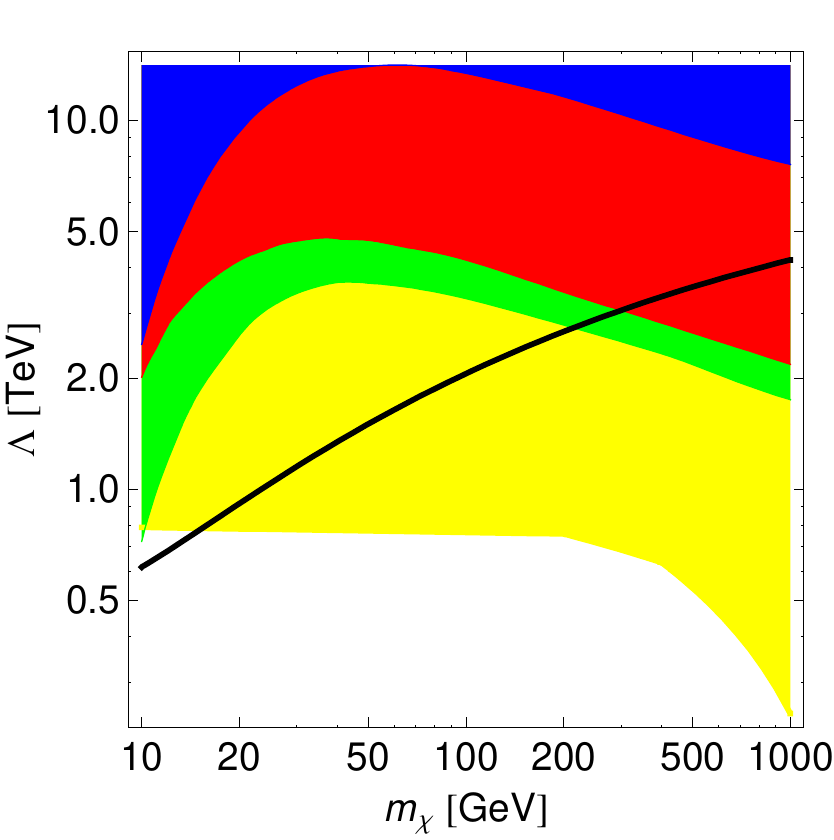}
    ~~~
    \includegraphics[width=0.45\textwidth]{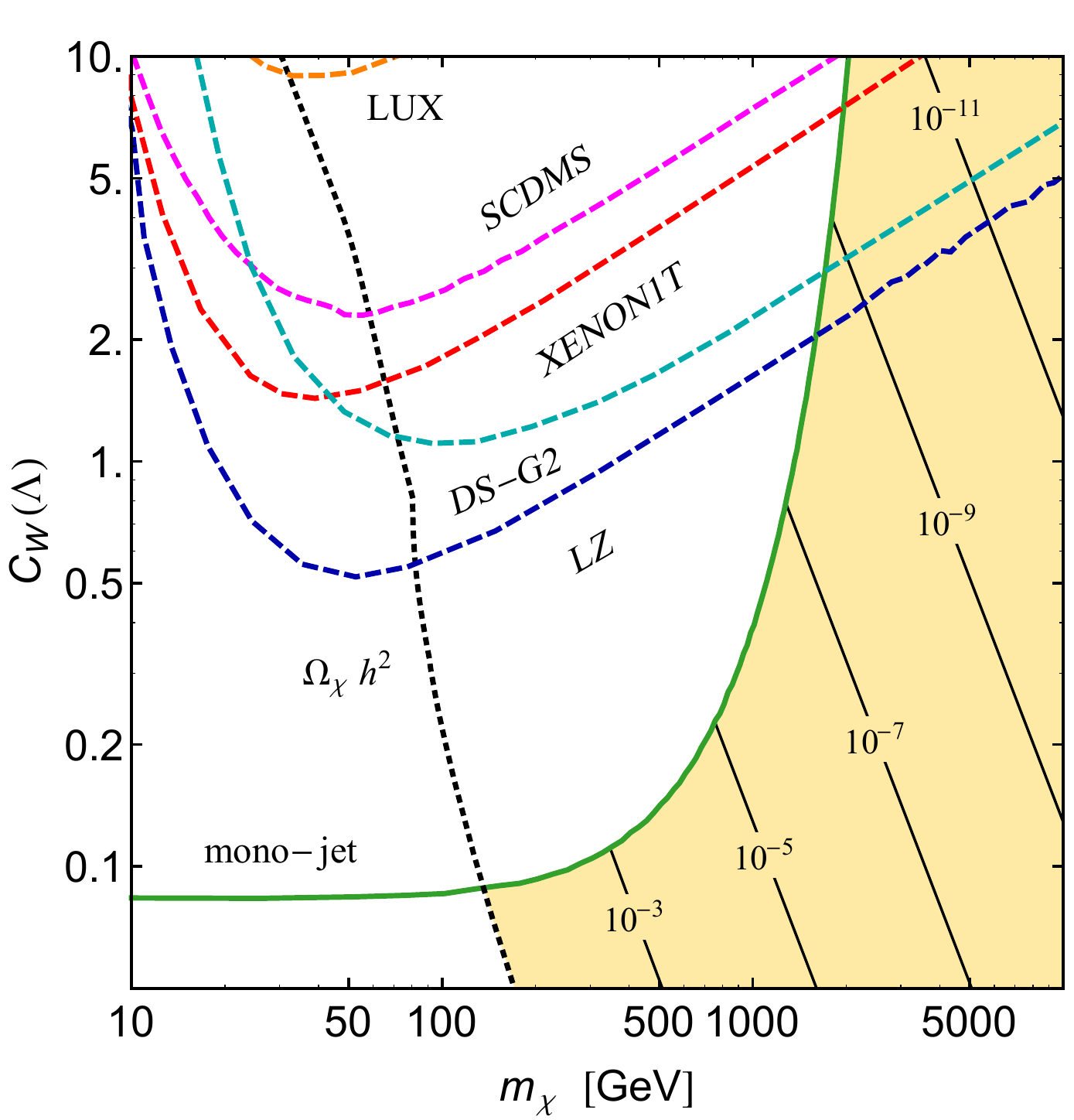}
\caption{Left: Allowed regions from LHC searches (yellow) and SI WIMP--nucleon scattering from LUX (green). Projected allowed regions for SCDMS (red) and XENON1T (blue) are also shown, as well as the curve giving the correct thermal relic density (black). Here we set $C^{VA}_{qq} = 1$ while all other Wilson coefficients are assumed to be zero.
Right: Restrictions in the $m_\chi$--$\hspace{0.25mm} C_W (\Lambda)$ plane, assuming DM to be Majorana and setting $\Lambda = 300 \, {\rm GeV}$. The green curves illustrate the best limits from missing ${E}_T$ searches at the LHC, while the black dotted lines correspond to the observed value $\Omega_\chi h^2 = 0.11$ of the relic density. The colored dashed curves mark the bounds from existing and future direct detections experiments. The currently allowed parameter regions are indicated by yellow shading. The contour lines denote the fraction of the observed relic density obtained from the operator under consideration.}\label{plot:EFT}
\end{figure}

At dim-6 the operator $O^{VA}_{qq}=\bar\chi {\gamma^\mu } \chi  \; \bar q \, {\gamma_\mu\gamma^5 q}$ mixes into $O_{HH D}^S = \bar \chi \, {\Gamma^\mu } \chi \, [{H^\dag }\overleftrightarrow{D}^\mu H]$ ($H$ being the SM Higgs doublet and $\overleftrightarrow{D}$ the covariant derivative) which then generates threshold corrections to $\bar\chi {\gamma^\mu } \chi  \; \bar q \, {\gamma_\mu}q$ entering spin independent direct detection~\cite{Crivellin:2014qxa}. The resulting bounds are shown in the left plot of Fig.~\ref{plot:EFT} depicting that even though the contribution is loop suppressed, direct detection gives stronger bound unless DM is very light.

At dimension dim-7 a similar effect occurs for the operators $O_W = \bar \chi \chi \, W_{\mu \nu }  W^{\mu \nu}$ involving electroweak field strength tensors. Again, this operator enters direct detection only via mixing and threshold correction~\cite{Crivellin:2014gpa}. The resulting bounds are shown in the right plot of Fig.~\ref{plot:EFT}. In this case the collider bounds are in general stronger~\cite{Crivellin:2015wva}, unless dark matter is quite heavy.

\section{MSSM}

\begin{figure}[t]
	\centering\includegraphics[scale=0.55]{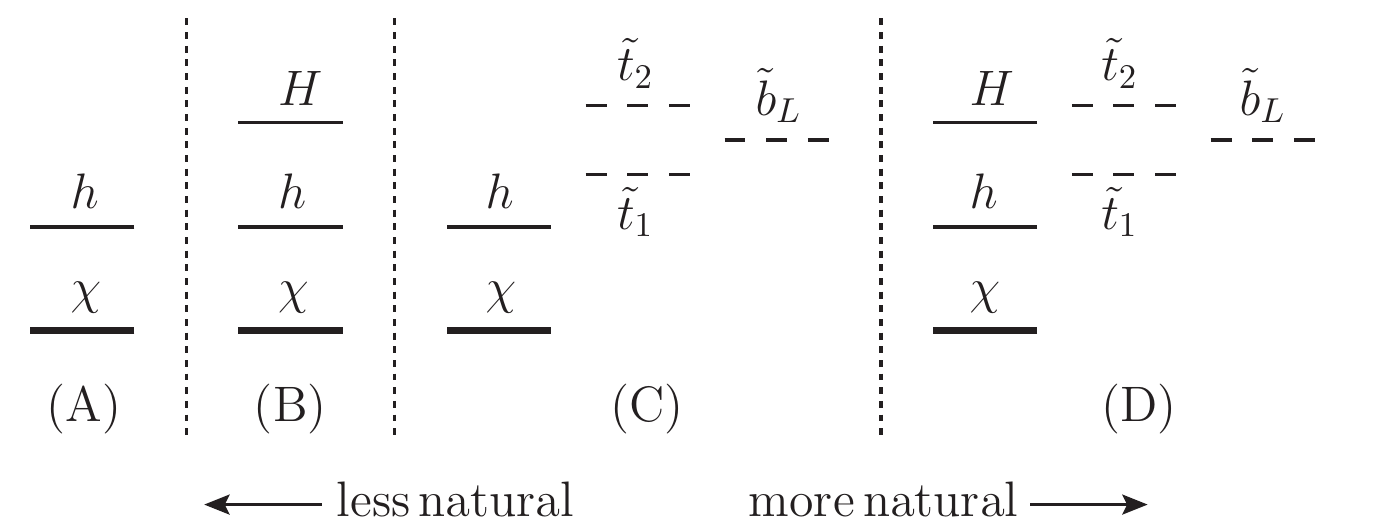}~~
		\centering\includegraphics[scale=0.55]{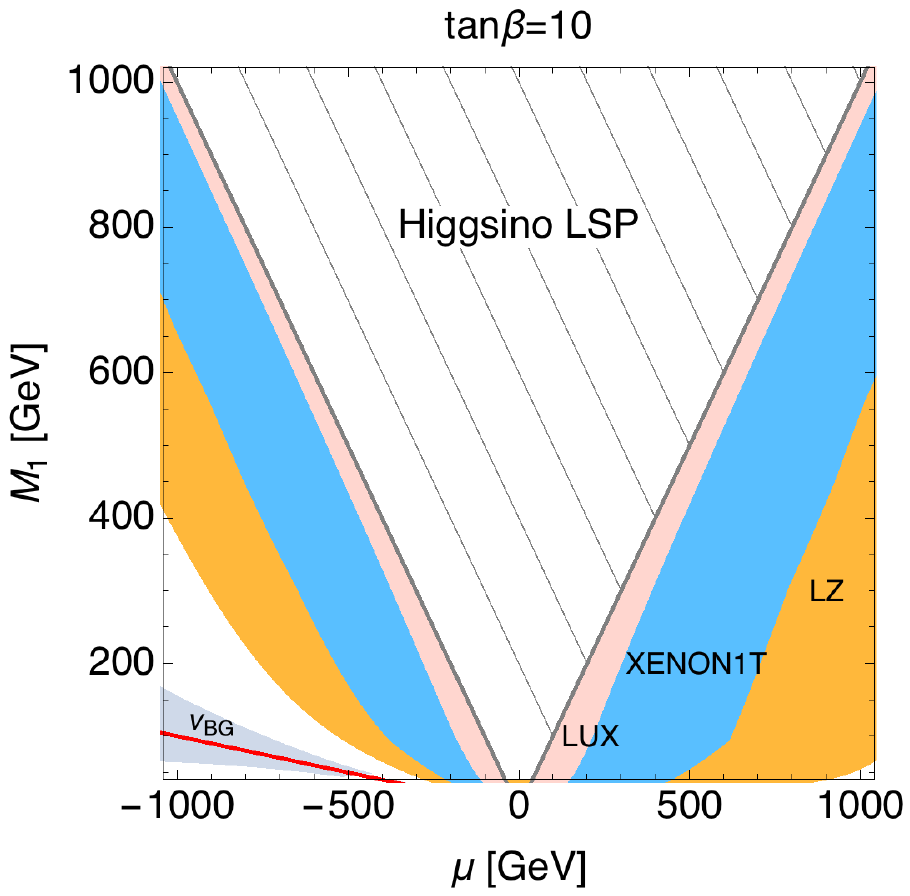}
	\caption{Left: Spectra of the simplified models for SI $\chi$--nucleus scattering considered in this work.  For each model, the SM-like Higgs is denoted by $h$, 
	while all other states are assumed to lie below 1 TeV, including Higgsinos (not shown).  From left-to-right, the spectra become increasingly more natural 
	as one includes the additional CP-even Higgs $H$ and third-generation squarks $\tilde{t}_1,\tilde{t}_2,\tilde{b}_L$. Right: Current and projected limits on SI $\chi$--xenon scattering due to $h$ exchange with $\tan\beta=10$. The pink band shows the existing constraints from LUX, while projected limits from XENON1T and LZ are given by the blue and orange regions respectively. The blind spot where the SI cross section vanishes is denoted by the red line and lies within the irreducible neutrino background ($\nu_\mathrm{BG}$) shown in gray.  The 
triangular, hatched region corresponds to the case where the LSP is Higgsino-like.}
	\label{fig:simple}
\end{figure}

Following Ref.~\cite{Crivellin:2015bva}, we use naturalness as a guiding principle in order to study neutralino dark matter scattering in the MSSM (see also Ref.~\cite{Barducci:2015ffa} for a recent analysis). In the left plot of Fig.~\ref{fig:simple} we show four simplified spectra which are increasingly natural ($A$ to $D$). Interestingly, in all scenarios blind spots \cite{Hisano:2012wm,Cheung:2012qy,Huang:2014xua,Anandakrishnan:2014fia} with vanishing scattering cross section can occur. In the proximity of these blind spots isospin violation is enhanced, making a precise determination of the scalar couplings to nucleons crucial~\cite{Crivellin:2013ipa}.\footnote{The same scalar couplings to the nucleon are also important for $\mu\to e$ conversion in nuclei~\cite{Crivellin:2014cta}.}

\begin{figure}[t]
	\centering
    \includegraphics[width=0.48\textwidth]{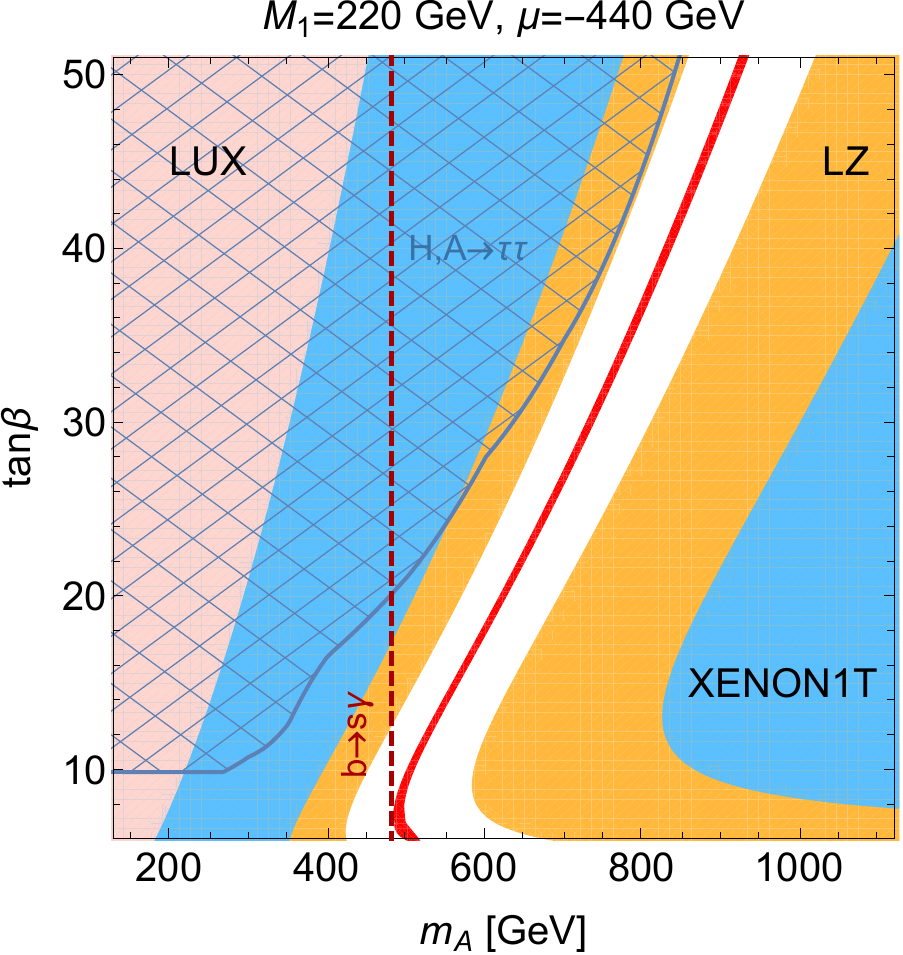}
    ~~~
    \includegraphics[width=0.48\textwidth]{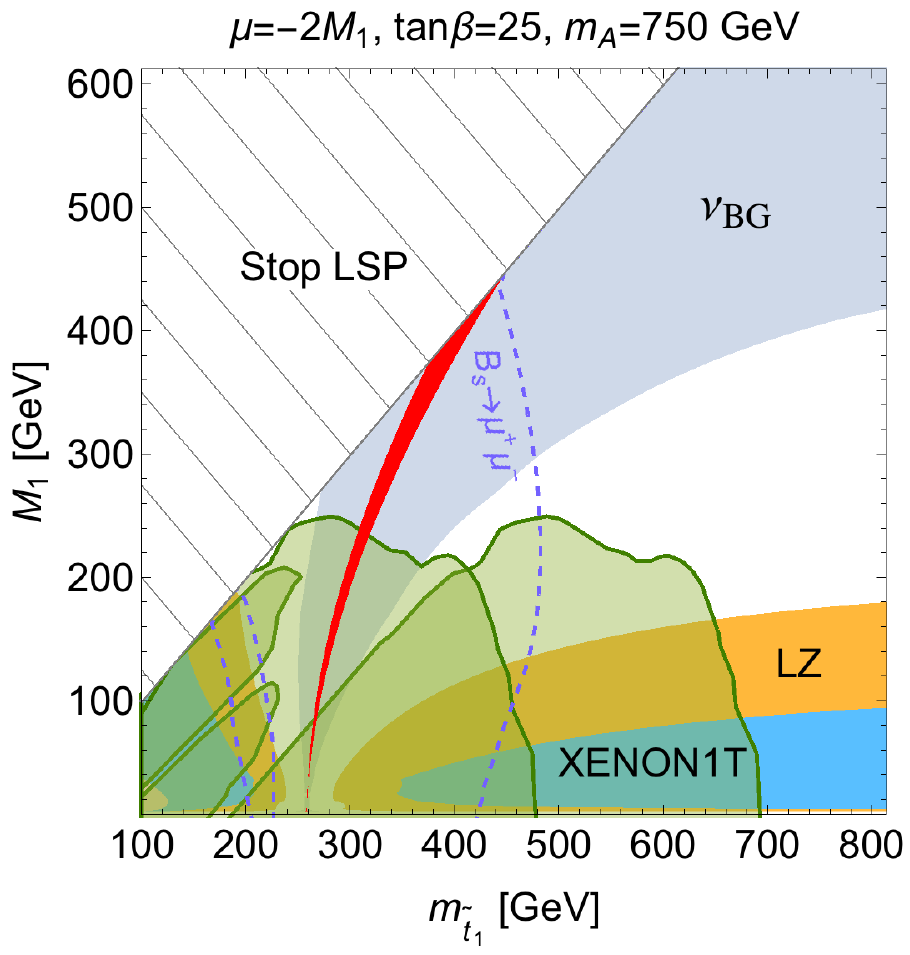}
\caption{Left: Current and projected limits on SI $\chi$--xenon scattering due to $h,H$ exchange with different benchmark values for $M_1$ and $\mu$. The cross-hatched region in dark-blue corresponding to CMS limits on $H,A\to \tau^+\tau^-$.  The region to the left of the dark-red dashed line at $m_A \cong m_{H^+} \simeq 480$ GeV is excluded by $b\to s\gamma$. \newline	Right: Current and projected limits in the $(m_{\tilde{t}_1},M_1)$ plane from $h,H$ and $\tilde{t}_{1,2},\tilde{b}_L$ exchange in $\chi$--xenon scattering. 	In the figures, the value of $m_A$ is increased for fixed $\tan\beta$.
}\label{plot:MSSM}
\end{figure}

In the case in which DM interactions are transmitted by the SM Higgs only, a blind spot occurs at $M_1 + \mu s_{2\beta} = 0$ as shown in the right plot of Fig.~\ref{fig:simple}. If we consider in addition the heavy CP-even Higgs $H^0$ (whose mass is nearly degenerate with the CP-odd Higgs $A^0$) the situation is more interesting, as we do not only have additional contributions to DM scattering but also get effects in $b\to s\gamma$ \cite{Misiak:2015xwa} and obtain bounds from LHC searches for $A^0\to\tau^+\tau^-$ \cite{CMS:2013hja} whose interplay is shown in the left plot in Fig.~\ref{plot:MSSM}. The occurrence of a blind spot where the $h^0$ and the $H^0$ contributions cancel is possible. Interestingly, future LHC searches for $A^0\to\tau^+\tau^-$ will be able to cover this region in parameter space which cannot be tested with direct detection.

The situation if in addition squarks of the third generation are included (the presence of a left-handed stop requires a left-handed sbottom as well due to $SU(2)_L$ gauge invariance) as dynamical degrees of freedom (scenario D) is shown in the right plot in Fig.~\ref{plot:MSSM}. Here the complementarity of LHC searches for stops and sbottoms with DM direct detection is illustrated as well as the effect in $B_s\to\mu^+\mu^-$ which we calculated with SUSY$\underline{\;\;}$FLAVOR~\cite{Crivellin:2012jv}. Again, part of the region in the proximity of the blind spot which cannot be covered by direct detection is already ruled out by LHC searches whose sensitivity to high masses will significantly increase at the $14\,$TeV run.

\section{Conclusions}

In these proceedings we reviewed the interplay between DM direct detection, flavor, and LHC searches by highlighting two prime examples: First we considered the EFT approach. Here LHC searches give complementary constraints on operators which enter spin independent scattering only at the loop level. Second, we considered the MSSM where LHC searches for stops, sbottoms, and heavy Higgses place constraints on the parameter space which are complementary to flavor observables and direct detection. We identify regions in parameter space with blind spots which cannot be covered by direct detection, but can be covered by LHC searches.

\section*{Acknowledgments}
We thank the organizers of the invitation to \emph{Moriond Gravitation} 2015 and for the opportunity to present these results. A.C. is supported by a Marie Curie Intra-European Fellowship of the European Community's 7th Framework Programme under contract number (PIEF-GA-2012-326948).

\bibliography{BIB}

\end{document}